\documentclass[runningheads]{llncs}
\usepackage{graphicx}
\usepackage{color}
\usepackage{subfigure}
\usepackage{url}
\usepackage[markup=underlined]{changes}

\begin{document}
\title{Child-Robot Interaction Studies During COVID-19 Pandemic} 
%
%
\author{Pinar Uluer\inst{1,2}\orcidID{0000-0003-2923-6220} \and
Hatice Kose\inst{2}\orcidID{0000-0003-4796-4766} \and
Agnieszka Landowska\inst{3}\orcidID{0000-0002-4728-689X} \and
Tatjana Zorcec\inst{4}\orcidID{0000-0002-4956-5163} \and
Ben Robins\inst{5}\orcidID{0000-0002-1646-901X}\and
Duygun Erol Barkana\inst{6}\orcidID{0000-0002-8929-0459}}
\authorrunning{P. Uluer et al.}
%
\institute{Galatasaray University, Turkey \and
Istanbul Technical University, Turkey \and
Gdańsk University of Technology, Poland \and
University Children’s Hospital, Skopje, Macedonia \and
University of Hertfordshire, UK \and
Yeditepe University, Turkey}
%
\maketitle              
\begin{abstract}
The coronavirus disease (COVID-19) pandemic affected our lives deeply, just like everyone else, the children also suffered from the restrictions due to COVID-19 affecting their education and social interactions with others, being restricted from play areas and schools for a long time. Although social robots provide a promising solution to support children in their education, healthcare and social interaction with others, the precautions due to COVID-19 also introduced new constraints in the social robotics research. In this paper, we will discuss the benefits and challenges encountered in child-robot interaction due to COVID-19 based on two user studies. The first study involves children with hearing disabilities, and Pepper humanoid robot to support their audiometry tests. 
The second study includes the child-sized humanoid robot Kaspar and interaction games with children with autism spectrum disorder (ASD). 

\end{abstract}

\section{Introduction}

In the times of pandemics, robots will be useful to handle the social aspect of the interaction which is lost during the outbreak by assisting humans to maintain social distancing, to monitor and improve mental health, and to support education~\cite{scassellati2020}. Feil-Seifer et al. present the current challenges faced by human-robot interaction (HRI) researchers in the time of pandemic and its impact on the research studies, research topics and student training; and they propose to use this new paradigm shift to reflect on the current research practices to design and implement new guidelines improving the field overall~\cite{feilseifer2020}.

The COVID-19 outbreak and its restrictions also impacted children, especially younger ones who benefit greatly from their physical interactions with their peers and the social learning through these interactions. 
These restrictions are especially a vital problem for children who need special rehabilitation or therapy. At this point, social robots and interaction games have the potential to solve the challenges introduced by these restrictions, since they can be personalized and adapted to meet the needs of children. 


In this paper, we present our own experiences in facing the challenge of COVID-19 in two user studies conducted with children and discuss the possible future directions in children-robot interaction (CRI) studies.

\section{CRI User Studies in Pandemics}

\subsection{Challenges}
The user studies in times of pandemic present major challenges for researchers working on in-person experience with social robots. In the two aforementioned CRI user studies, we are working with hearing-impaired children having hearing aid or cochlear implant to improve their engagement during their auditory rehabilitation (RoboRehab), and children with autism to help them improve their social skills and emotional representations (EMBOA). 
The challenges we faced during these two user studies may be grouped in three categories: 

\textbf{(1)} Organizational aspect: The organizational issues consist of the governmental/institutional regulations for public health to prevent the spread of COVID-19. 
All in-person CRI procedures go through an ethical committee to protect the rights and welfare of the participants before the recruitment process starts. The procedures during the pandemic are updated recurrently based on the new scientific findings and are presented to ethical committee and checked for the governmental regulations provided by the local health administrations. 

As expected, all the studies with children came to an abrupt stop when the pandemic started and the lockdown happened in several countries. When the restrictions have been eased and the clinics and schools for children have started to operate again, to keep the safety of children and their personnel, they are not motivated to continue the studies to avoid any additional risk factors even though the studies are 
approved for the health and safety considerations. The therapists are also 
intimidated by the additional interactions with children and their parents during the pandemics. On the other hand, recruitment process presents another challenge since the parents are not enthusiastic about their children to interact with a robot/clinician/therapist or researcher interacting with other children. Therefore some of the planned studies in various countries are canceled or postponed indefinitely. 


\textbf{(2)} Technical aspect: The technical challenges issued by COVID-19 are mostly based on the sanitization of the equipment (i.e. the robot itself, sensors, surfaces) and the challenges caused by the selected sensor technologies. 

The sanitization of the equipment is a laborious process because each one of them has a specific procedure for sanitization. Even though most of the equipment may be sanitized by exposure to a UV Lamp (E4 wristband, microphones, test surfaces, etc) or by isopropyl alcohol-based disinfectants (Pepper robot), the process takes longer and prevents the interaction flow of the studies with group of participants therefore the studies are rescheduled as one-to-one interaction scenarios without any tactile interaction. 

\begin{figure}[thbp]
  \centering
  \subfigure[]{\includegraphics[width=0.17\textwidth]{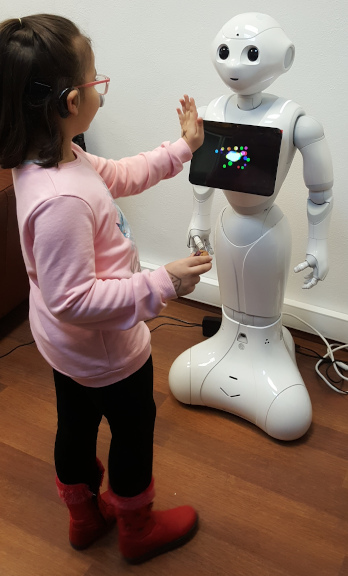}\label{fig:pepper}}
  \subfigure[]{\includegraphics[width=0.375\textwidth]{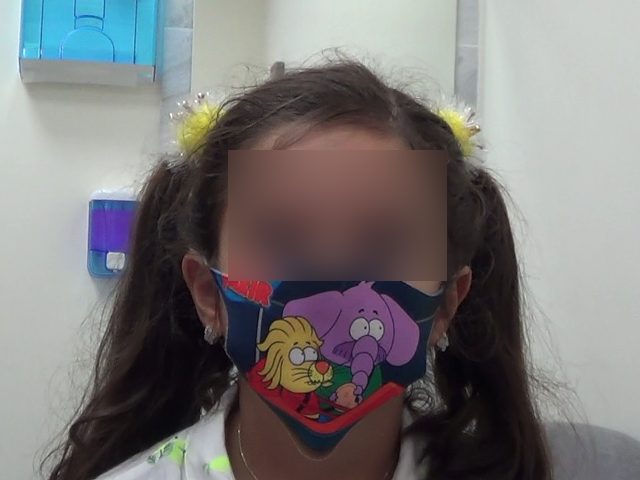}\label{fig:roborehab}}
  \subfigure[]{\includegraphics[width=0.375\textwidth]{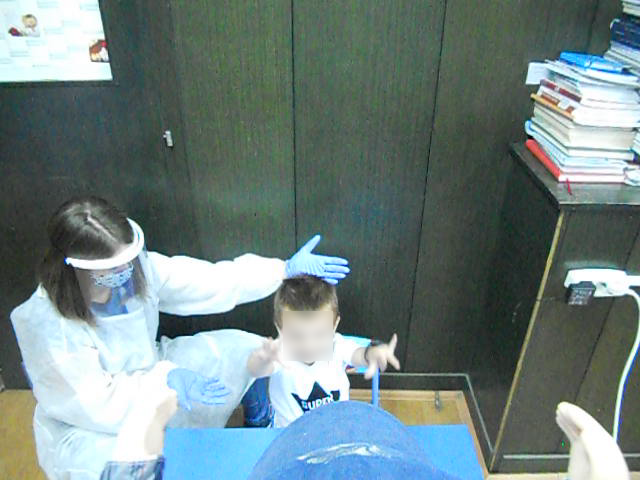}\label{fig:emboa}}
  \caption{(a) A screenshot from RoboRehab study before COVID-19 pandemic, (b) Participant with mask (RoboRehab), (c) Participant without mask (EMBOA) }
  \label{fig:cri-study}
\end{figure}

The emotion or stress detection from facial expressions captured by video cameras, which are fortunately non-contact sensors, is already a challenging task with children's non-posed real-time interaction data and now with the pandemic there is also the additional factor of the facial mask. In some cases with younger children, it is possible to isolate the interaction area and provide a mask-free interaction (Fig.~\ref{fig:emboa}) but in other cases with older children it is compulsory to wear the mask, especially in clinics (Fig.~\ref{fig:roborehab}). The masks not only influence the analysis of facial expressions, but also have an effect on the audio recordings captured during the interaction. 

In order to overcome the time delay caused by the sanitization process, the use of see through barriers such as plexiglass covers may seem to provide an alternative solution but to introduce such materials between the child and the robot or sensor devices may also alter the audio (utterance and vocalization of children) and video (eye gaze detection) recordings, and it may pose other obstruction or illumination challenges to process the interaction data.

\textbf{(3)} Interaction-related aspect: The preventive measures have a major impact on the design of the interaction. All the interaction scenarios are revised and the scenarios encouraging natural tactile interaction are discarded, hence, loss of modality. The tactile interaction may not play a crucial role in most HRI studies but the tactile perception is essential for children with autism~\cite{caldwell1996}. The sanitization of the robot and the child's hands may be a solution to introduce the tactile experience into the study, but it is not a feasible solution, because Kaspar robot is covered by a silicone skin and a set of baby clothes requiring multiple level sanitization process. And it will be hard to sanitize the robot every time the children put their hands to its mouth or nose during the interaction.

Another factor is the protective clothing worn by the experimenters, the whole setup with the coat, gloves, mask and visor may have a negative effect on the children and reduce the interaction gain. 

\subsection{User Studies}
After the outbreak, we conducted pilot studies for both projects to investigate and explore the experimental setup and procedure according to these challenges.

The first user study was part of RoboRehab project~\cite{website-roborehab}, we are working on an affective and social robotic assistant for the audiology rehabilitation of hearing-impaired children to assist them during their clinical consultations. This interdisciplinary project involves emotion recognition; game-based auditory test production; and interaction with a social humanoid robot, Pepper. We aim to use machine learning-based emotion and stress recognition approaches to understand the children's feelings, decide the assistive robot's actions, and design a feedback mechanism to improve the performance and reduce the stress of children during their audiometry tests. 
Six children (6-7 years old) with cochlear implant participated in the study for RoboRehab where we tested the tablet-based gamification setup of an auditory test. The consultation room was aired and the experimenters wore full protective clothing. The children were asked to put a mask and wash their hands. After the tablet game was finished; the tablet, the smart wristband and all the physical test materials were sanitized and the iteration continued with the airing of the room.  

The second user study was part of EMBOA project~\cite{website-emboa}, an international research project combining three domains of research; autism therapy, social robots (Kaspar) and automatic emotion recognition, in order to develop guidelines and practical evaluation of applying emotion recognition technologies in robot-supported intervention in children with autism. The motivation is to investigate the feasibility of the available practices and find a novel approach for creating an affective loop in child-robot interaction that would enhance the intervention regarding emotional intelligence building in children with autism. 
The study for EMBOA project was conducted with 3 children with ASD. After the organizational issues were resolved, an isolated room was reserved for the participants and their families, for them not to be subjected to additional interaction with clinicians or research personnel. The study day started early with the natural ventilation of the room for 20 minutes combined with an air purifier. In the mean time, the physical setup was prepared and the experimenters put on their protective clothes (coat, gloves, mask and visor). When the child arrived 
the temperature check was done, and the child's hands and shoes were disinfected. During the session the child was not allowed to touch Kaspar. 
At the end of the session, the physical setup (table, chairs, assistive toys, etc.) and the compliant electronic materials (smart wristband, microphones,  etc.) were sanitized. And the experimenters had a change of protective clothing and ventilate the room another 20 minutes, and the cycle continued.

\section{Conclusion}
The use of social robots will support the lives of children greatly in the pandemic times, where their social activities, education, therapies and even healthcare is restricted due to COVID-19. 
To implement the physical components of therapeutic interaction, such as touching the robots, becomes challenging due to these restrictions.
It is possible to put barriers between the child and robot and not let the child touch the robot, however this will restrict the natural interaction between the pair and
prevent the use of many therapy games involving touch. Likewise, at all times putting on masks is an essential rule especially in closed locations with several people, yet it occludes the facial features of children which are important for the robot to infer the emotions and mood of children. 
We can still use sensors such as E4 wristband to detect physiological signals, cameras such as depth cameras to detect actions and posture, and eye-trackers to detect attention without any problem under these pandemic circumstances. Due to COVID-19, the researchers in CRI field need to plan and consider the discussed issues for future setup for their studies.

\section*{Acknowledgments}
This study was co-funded by Erasmus Plus project of European Commission:
EMBOA, Affective loop in Socially Assistive Robotics as an intervention
tool for children with autism, contract no 2019-1-PL01- KA203-065096. This
publication reflects the views only of the author, and the Commission
cannot be held responsible for any use which may be made of the
information contained therein.

This study was partially supported by TUBITAK, RoboRehab project under contract no 118E214.

%
%
%
\bibliographystyle{splncs04}
\bibliography{references.bib}
%




\end{document}